\documentclass[prb,preprint,superscriptaddress,showpacs,floatfix]{revtex4-1}

\usepackage{graphicx}
\usepackage[graphicx]{realboxes}
\usepackage{amsmath}
\usepackage[english]{babel}
\usepackage{enumerate}
\usepackage{float}
\usepackage[update,prepend]{epstopdf}

\begin{document}


\bibliographystyle{apsrev4-1}

\title{Magnetic vortices as localized mesoscopic domain wall pinning sites}

\author{R. L. Novak}
\email[Corresponding author: ]{rafael.novak@ufsc.br}
\affiliation{Universidade Federal de Santa Catarina -- Campus Blumenau, Rua Pomerode, 710, 89065-300  Blumenau (SC), Brazil}
\author{L. C. Sampaio}
\affiliation{Centro Brasileiro de Pesquisas F\'isicas -- Rua Dr. Xavier Sigaud 150, 22290-180 Rio de Janeiro (RJ), Brazil}

\begin{abstract}

We report on the controllable pinning of domain walls in stripes with perpendicular magnetic anisotropy by magnetostatic coupling to magnetic vortices in disks located above the stripe. Pinning mechanisms and depinning fields are reported. This novel pinning strategy, which can be realized by current nanofabrication techniques, opens up new possibilities for the non-destructive control of domain wall mobility in domain wall based spintronic devices.

\end{abstract}


\maketitle

\section{Introduction \label{intro}}

Magnetic domain wall-based microelectronic devices are promising candidates for future data storage and spintronic technologies \cite{parkin, allwood_logic, allwood_review}. In these devices, the motion and careful positioning of domain walls (DWs) in thin film-based patterned sub-micron stripes underlie the functionalities of the device, and the precise control of DW motion evidently becomes one of the most important goals in device design \cite{parkin, allwood_logic}. The most common approaches rely either on localized modifications of material parameters such as the anisotropy constant, which can be modified by ion bombardment \cite{vieu, repain, franken_anisotropy}, or on structural modifications of the stripe such as holes \cite{adeyeye}, lateral extensions along the stripe \cite{cowburn_comb} or notches cut into its long edge \cite{parkin, hayashi_notch, hayashi_notch2, atkinson_notch, brandao_novak}. In all these cases, the goal is to induce strong DW pinning at certain positions. Recently, even standing acoustic waves have been proposed as pinning sites \cite{sound_idea} along magnetic stripes.

All these approaches rely on structural modifications that lead to different local distributions of demagnetizing fields that change either the internal DW structure and dynamics, or the local fields acting on the DW itself. Although very effective, these alternatives are destructive, since they introduce modifications on the medium where DW motion is occurring. As this can lead to undesired side effects on the magnetization processes, non-destructive methods, based on the magnetostatic coupling between DWs and magnetic structures external to the medium, offer an attractive alternative way to control the DW positioning through pinning.

One approach to non-destructive DW pinning is the introduction of localized magnetic fields generated by overlying nanomagnet arrays \cite{dots1, dots2} separated from the stripe by non-magnetic spacer layers thick enough to ensure that the coupling is primarily magnetostatic. The asymmetric pinning generated by these nanomagnets provides a way to locally pin DWs in a controllable and potentially reprogrammable fashion. This method, furthermore, introduces a new degree of freedom to control the pinning of DWs because different DW mobilities are possible by controlling the relative alignment of the applied field and the orientation of the magnetization of the nanomagnets \cite{dots1}. Even though effective, this method relies on large $(50 \, \mathrm{x}\, 50 \mu\mathrm{m}^2)$ arrays, which are evidently not suitable for very small/narrow stripes. In this case, single nanomagnets acting as pinning sites are desired, but their fabrication still presents a challenge. Recently, Franken et al. \cite{franken} succeeded in growing a single magnetic nanopillar on top of a stripe and showed that the perpendicular components of its stray fields effectively work as a source of pinning for a magnetic DW moving along an underlying stripe with perpendicular magnetic anisotropy (PMA), and that the pinning can be tuned by the height of the pillar as well as its magnetic state. In another work, van Mourik et al. \cite{van_mourik} demonstrated the feasibility of using a single overlying nanomagnet with in-plane magnetization as source of switchable pinning of DWs in magnetic stripes with PMA. As in Franken et al.\cite{franken}, the magnetic state of the nanomagnet, and the thickness of the spacer layer, determine the pinning strength at the site. Nanomagnets, however, may present many ground states that depend on the material parameters and geometry, and consequently maintaining a nanomagnet in single domain state may impose restrictions to device geometry. In many nanomagnet geometries, the ground state exhibits a magnetic vortex \cite{cowburn, shinjo, wachowiak, hubert_book, feldtkeller} characterized by an in-plane magnetization that curls around the center of the nanomagnet and a strong and spatially localized out-of-plane component, known as the vortex core (VC), at its center. The combination of these in-plane systems with PMA films in a multilayer device could lead to new functionalities, because the strong stray fields emanating from the VCs could act as sources of localized, non-destructive pinning sites for DWs in, for example, an underlying stripe with PMA. Further pinning could be achieved by coupling of the in-plane components of the vortex magnetization and of the DW stray field, effectively presenting two sources of DW pinning in these hybrid in-plane/out-of-plane magnetic systems. Similar systems have already been investigated in extended PMA films coupled to a single Permalloy nanomagnet \cite{heldt, wohlhuter} with a vortex ground state. In these studies, the coupling between the VC and the underlying out-of-plane domain structure was demonstrated, but since there was no spacer layer separating the different materials, both magnetostatic and exchange interactions contributed to the observed coupling. The authors did not investigate a scenario where only the magnetostatic fields contributed to the coupling, as was done in \cite{dots1} and \cite{dots2}. The understanding of how the magnetostatic or the exchange interaction contribute individually to the coupling between DWs and vortices is fundamental to further developments of these hybrid systems, especially since magnetostatic fields play a major role in the statics and dynamics of magnetic nanostructures.

In this work, we report the results of micromagnetic simulations demonstrating the feasibility of using a soft magnetic disk-shaped nanomagnet with in-plane magnetization and a vortex ground state as a source of purely magnetostatic pinning for Bloch DWs in underlying magnetic stripes with PMA. The purely magnetostatic coupling between the vortex and the DW gives rise to strong and asymmetric pinning which could be exploited in spintronic devices. Simulated hysteresis loops show that the pinning involves coupling of both the in-plane and out-of-plane components of the DW and vortex magnetizations, indicating a complex pinning scenario, arising purely from magnetostatic interactions, highlighting the major role this interaction plays in pinning site engineering in magnetic DW-based devices.

\section{Micromagnetic simulations \label{sims}}

The simulations were performed with the Mumax3 package \cite{mumax3}. The simulated system consisted of a $2000$ nm long, $512$ nm wide Co-like stripe with $M_s = 1135$ kA/m, $2$ nm thickness, exchange stiffness $A_{ex} = 17$ pJ/m, damping constant $\alpha = 0.5$ and perpendicular uniaxial anisotropy constant $K_u = 1240$ kJ/m$^3$. This stripe was capped by a variable thickness empty ``spacer layer''. On top of this layer, a $512$ nm wide and $24$ nm thick NiFe-like disk was put in the midpoint of the Co stripe (Fig. \ref{fig:cycles}, inset). The disk has the following material parameters: $M_s = 796$ kA/m, $A_{ex} = 13$ pJ/m, $\alpha = 0.5$ (for faster relaxation) and no intrinsic magnetic anisotropy. These regions are divided in $3.9$ x $4$ x $2$ nm$^3$ cells. The stripe is always initialized with a Bloch DW near the left end of the stripe, separating an ``up'' domain ($m_{z}^{stripe} = +1$) on the left from a ``down'' domain on the right ($m_{z}^{stripe} = -1$). This domain structure will be present in the stipe in all simulations. The disk is initialized in a vortex state with definite circulation ($c = +1$ for counterclockwise and $-1$ for clockwise circulation) and core polarity ($p = +1$ for and ``up'' core magnetization and $-1$ for a ``down'' core magnetization). The system is then allowed to relax using Mumax’s relaxation routine (``minimize'') which uses a conjugate gradient method to evolve the magnetization until the ground state configuration is reached. Following each relaxation step, a magnetic field is applied along the $z$ axis (perpendicular to the stripe plane) in $10$ Oe steps, driving the magnetization reversal of the stripe through DW displacement. This way, hysteresis loops of the stripe can be obtained. The perpendicular field does not significantly affect the magnetic state of the disk, which stays in its initial vortex state until interacting with the DW. Time-driven, torque minimization dynamical simulations were also performed, yielding the same hysteresis loops as the relaxation routine outlined above.

\section{Results and discussion \label{res}}

\begin{figure}
\centering
\includegraphics[width=0.75\textwidth]{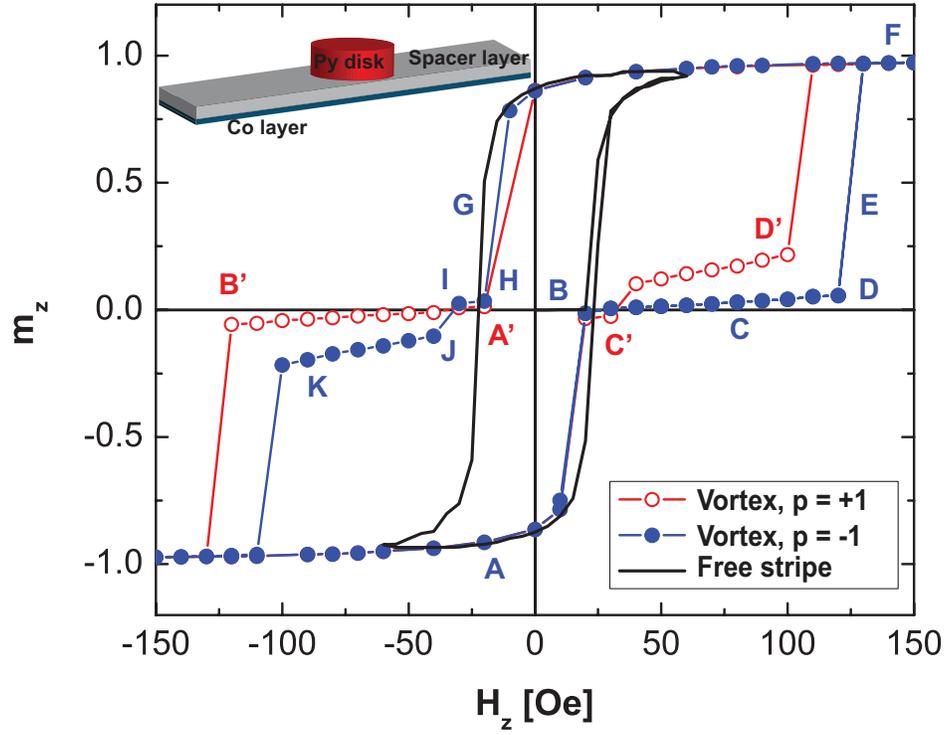}
\caption{Hysteresis loops of the Co stripe coupled to a vortex with $p = +1$ (red symbols and line) and $p = -1$ (blue symbols and line). The spacer layer is $6$ nm thick. The letters indicate points along the loops that are discussed in the text and depicted in Fig. \ref{fig:snapshots}. The black continuous line corresponds to a hysteresis loop of a free Co stripe. The inset shows the geometry of the system investigated. (Color online). \label{fig:cycles}}
\end{figure}

\begin{figure}
\centering
\includegraphics[width=0.6\textwidth]{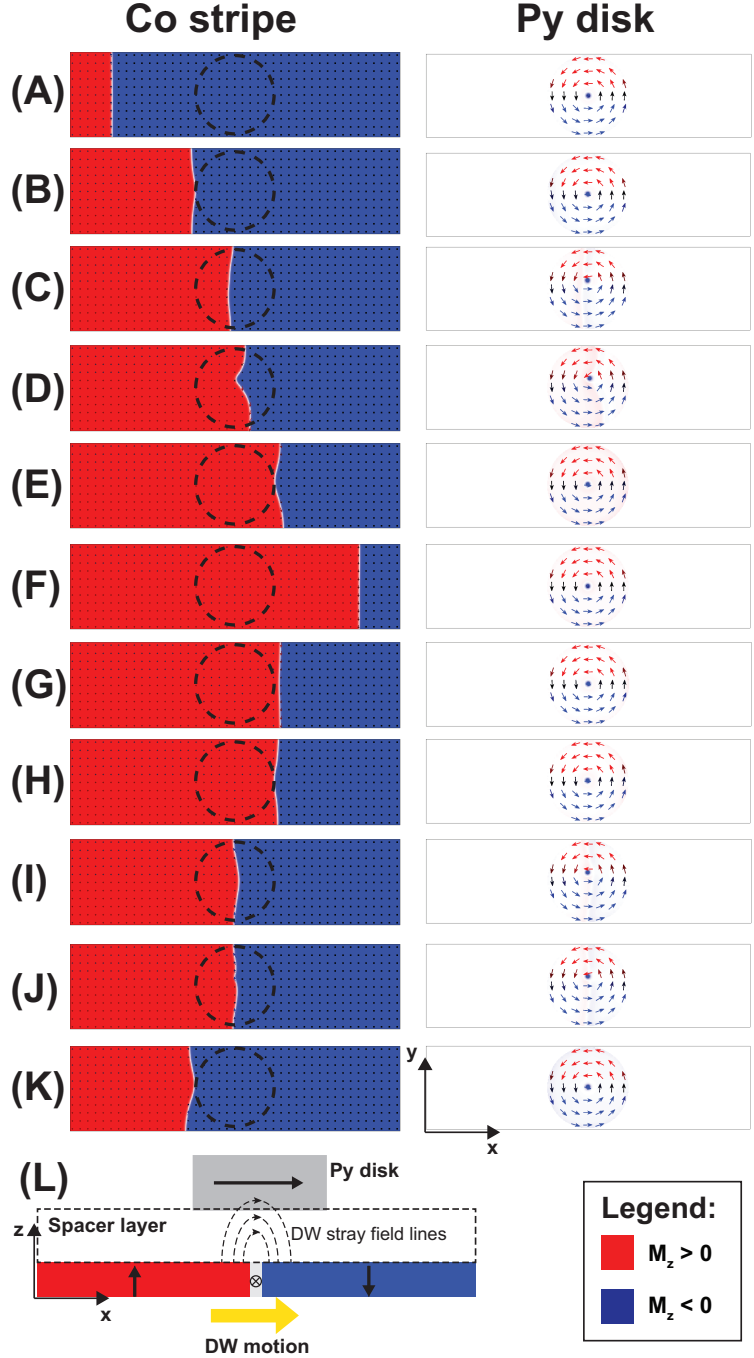}
\caption{(A)-(K) Snapshots of the Co stripe (left column) and the Py disk (right column) during the hysteresis loop simulation shown in Fig. \ref{fig:cycles}. The dashed circles indicate the position of the disk. The out-of-plane component of the reduced magnetization ($m_z$) is shown in red ($m_z > 0$) and blue ($m_z < 0$). The in-plane magnetization of the disk is represented by the arrows. (L) Lateral view of the DW during the stripe magnetization reversal. (Color online). \label{fig:snapshots}}
\end{figure}

We will first consider the cases of a disk with $c = +1$ and $p = +1$ or $-1$, and a $6$ nm thick spacer layer. Hysteresis loops of the Co stripe obtained in these two cases are shown in Fig. \ref{fig:cycles} (blue and red symbols) and snapshots of the magnetization state of the stripe and the Permalloy disk corresponding to selected points along the loop are shown in Fig. \ref{fig:snapshots}. The loops have a different shape when compared to a free Co stripe with the same geometry (black line), indicating the complex magnetization reversal process taking place.

In the case of negative core polarity ($p = -1$, blue circles and curve) the magnetization process is characterized by the ``free'' propagation of a DW under an external magnetic field $H_z = +20$ Oe, starting close to the left edge of the stripe (Figs. \ref{fig:cycles} and \ref{fig:snapshots}, A) and moving towards the right until it reaches the edge of the disk (Figs. \ref{fig:cycles} and \ref{fig:snapshots}, B). At this point, the DW is ``pulled'' towards the region underneath the disk and continues to move slowly towards the disk center, causing the VC to move upwards (Figs. \ref{fig:cycles} and \ref{fig:snapshots}, C). This happens because of the strong in-plane component of the DW stray field acting along the $+x$ direction (Fig. \ref{fig:snapshots}, L). Since the vortex has positive circulation ($c = +1$), the application of an in-plane field along $+x$ pushes the VC upwards, as the results clearly show. Eventually, the DW and the VC both reach equilibrium positions (Figs. \ref{fig:cycles} and \ref{fig:snapshots}, C) under $H_z = +20$ Oe. Then, $H_z$ is increased in $10$ Oe steps, causing the DW to move further to the right (Figs 1 and 2, B - D). As it approaches the VC, the DW starts to display an increasing bowing around it, caused by magnetostatic coupling to out-of-plane components of the VC stray field (Figs. \ref{fig:cycles} and \ref{fig:snapshots}, D). This coupling is the main source of this localized DW pinning and the DW bowing, since due to the large width of the stripe ($512$ nm) compared to the lateral dimensions of the VC ($\sim 20$ nm), DW sections far from the VC position will tend to move away under the applied $H_z$, while the DW section close to the VC will stay strongly pinned at its position, causing the significant bowing observed. Further increasing $H_z$ causes the bowing to get stronger and the core to move further up. When $H_z$ reaches $130$ Oe, the DW depinning is observed (Figs. \ref{fig:cycles} and \ref{fig:snapshots}, D -- E). The DW moves past the VC and quickly jumps to the right edge of the disk (Figs. \ref{fig:cycles} and \ref{fig:snapshots}, E) and eventually reaches the right edge of the stripe, leading to its magnetic saturation (Figs 1 and 2, F). Thus, the first half of the stripe hysteresis loop is obtained.


Now, is a negative field is applied ($H_z < 0$), the DW sense of motion will be reversed. Under $H_z = -20$ Oe, the DW will move from the right edge of the stripe towards the right edge of the disk (Figs. \ref{fig:cycles} and \ref{fig:snapshots}, F -- H). The motion is similar to what has been previously observed: upon reaching the disk edge, the DW slowly moves underneath it, while the VC moves up under the effect of the DW in-plane stray field. However, as the DW approaches the VC position (Figs. \ref{fig:cycles} and \ref{fig:snapshots}, I), it jumps towards its left side without any additional increase in applied field (Figs. \ref{fig:cycles} and \ref{fig:snapshots}, J), reaching its equilibrium position on the left side of the VC. Similar situations were reported in \cite{heldt, wohlhuter}, where it was shown that a VC tends to stay in equilibrium near DWs in an underlying magnetic film. Given the domain structure in the stripe, and the negative polarity of the VC, the magnetization of the propagating domain is parallel to the VC magnetization, effectively making it easy for the DW to cross the VC position at a relatively small applied field strength ($H_z = -40$ Oe). The opposite phenomenon is observed when $H_z > 0$ (``A'' and ``B'' points in Fig. \ref{fig:cycles}): since the vortex core magnetization, pointing ``down'' ($p = -1$), is antiparallel to the propagating domain magnetization, the core acts as an effective strong pinning barrier for DW propagation, causing the strong DW bowing observed (Fig. \ref{fig:snapshots}, D) and the high applied field necessary to precipitate the depinning process ($+130$ Oe). Finally, increasing the applied field value from $-40$ Oe to $-100$ Oe causes the DW to drag towards the left edge of the disk (Figs. \ref{fig:cycles} and \ref{fig:snapshots}, J -- K), again with its in-plane stray field component coupled to the in-plane magnetization of the disk (Fig. \ref{fig:snapshots}, L). The DW depins from the left edge of the disk at $-100$ Oe (Figs. \ref{fig:cycles} and \ref{fig:snapshots}, K), quickly moving towards the left edge of the stripe, leading to its magnetic saturation. It is important to notice that no VC reversal is induced by interaction with the underlying DW at any point of the hysteresis loop thus obtained.

If the VC polarity were positive ($p = +1$), with the circulation still positive ($c = +1$), the simulated hysteresis loop (Fig. \ref{fig:cycles}, red circles and curve) will be symmetric to this one, with the strong DW-VC pinning occurring at negative fields (depinning at $-130$ Oe, Fig. \ref{fig:cycles}, B’) and the weak pinning occurring at the positive field region of the loop (depinning at $+40$ Oe, Fig. \ref{fig:cycles}, C’). Again, the DW drags along the regions underneath the disk, depinning from its edges with the same applied fields, regardless of vortex core polarity.

\begin{figure}
\centering
\includegraphics[width=0.6\textwidth]{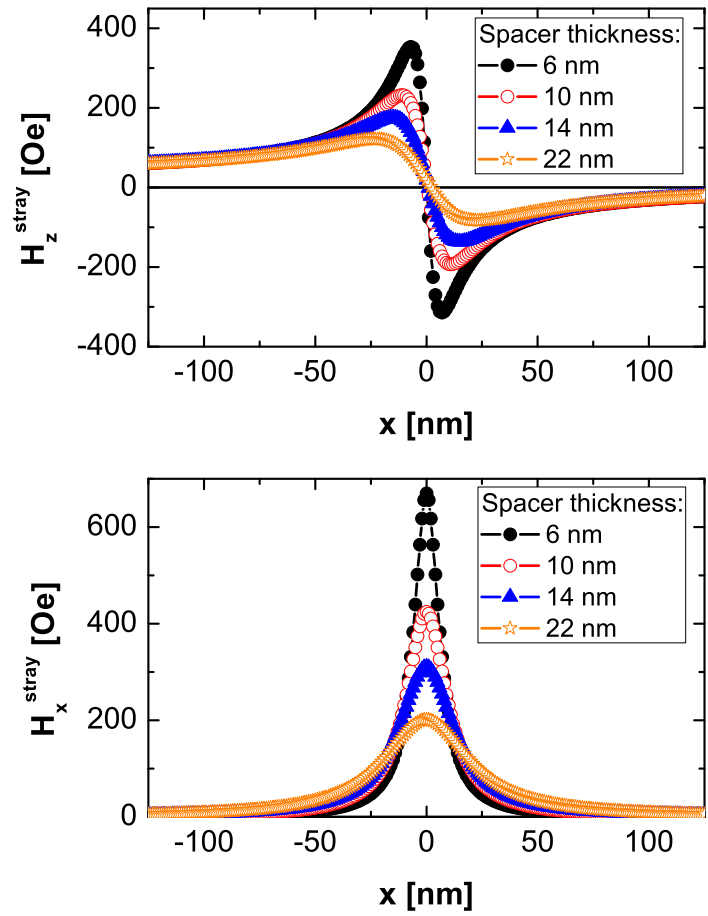}
\caption{Magnetostatic stray field profiles of a Bloch DW. (a) Out-of-plane component. (b) In-plane component along stripe length. (Color online). \label{fig:stray}}
\end{figure}

These results indicate that two coupling mechanisms between the DW and the vortex in the disk are present and contribute to the observed behavior: \emph{(i)} the coupling between the out-of-plane components of the DW stray field (Fig. \ref{fig:stray}) and the out-of-plane VC magnetization; \emph{(ii)} the coupling between the in-plane components of the DW stray field (Fig. \ref{fig:stray}) and the in-plane components of the vortex magnetization.

\begin{figure}
\centering
\includegraphics[width=0.75\textwidth]{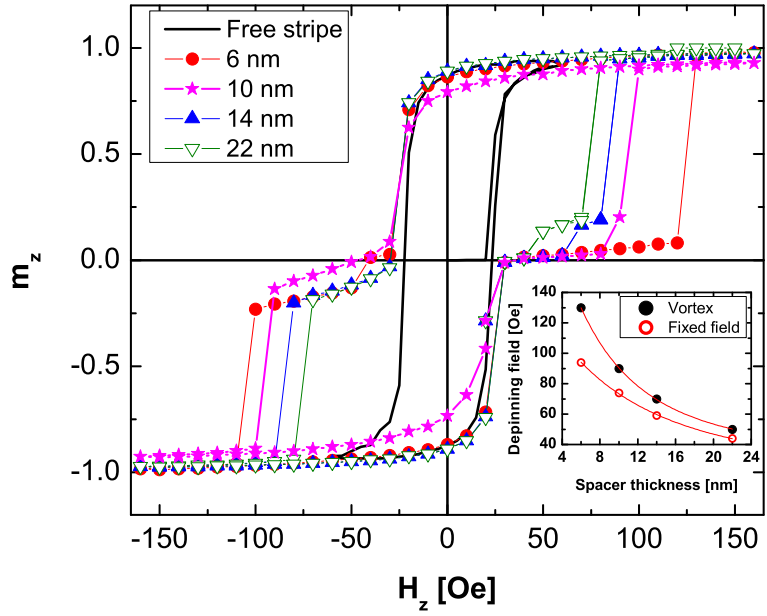}
\caption{Simulated hysteresis loops of the Co stripe coupled to the Py disk for several thicknesses of the spacer layer. The vortex has $c = +1$ and $p = -1$. The inset shows the fields where depinning of the DW from the vortex core is observed (filled circles) together with depinning fields obtained from simulations where only a fixed external out-of-plane field, mimicking the vortex core field, was applied to the Co stripe (open circles). (Color online). \label{fig:thicknesses}}
\end{figure}

Since the couplings leading to the observed DW asymmetric pinning are magnetostatic, increasing the distance between the stripe and the disk should decrease the coupling strength in both cases. This coupling strength can be inferred from the strength of the depinning fields extracted from the hysteresis loops. Simulated hysteresis loops with increasingly thicker spacer layers, but the same domain structure in the stripe and in the disk, are shown in Fig. \ref{fig:thicknesses}. For $10$ nm, $14$ nm and $22$ nm thick spacers, the hysteresis loops obtained are always similar to the one obtained with a $6$ nm spacer (Figs. \ref{fig:cycles} and \ref{fig:snapshots}), with the DW coupling to both the VC and to the in-plane components of the disk magnetization. On the other hand, the coupling strength decreases as we increase the separation between the stripe and the disk, as the depinning fields from the VC position (right side, Fig. \ref{fig:thicknesses}) and from the disk left edge (left side, Fig. \ref{fig:thicknesses}) show. The VC depinning fields are plotted against the spacer thickness in the inset. The strength of this depinning field decreases as an inverse cube law with the spacer thickness, further evidencing the dipolar magnetostatic nature of the coupling.

\begin{figure}
\centering
\includegraphics[width=0.75\textwidth]{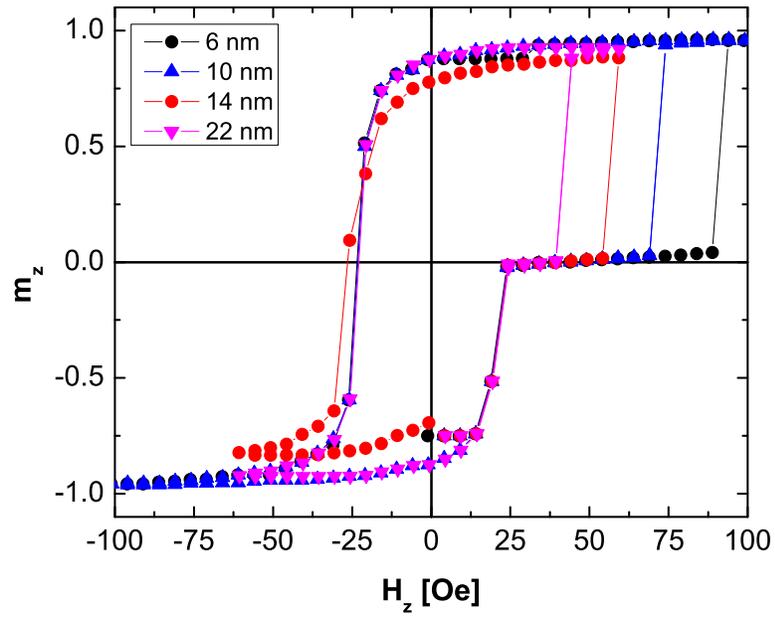}
\caption{Simulated hysteresis loops of the Co stripe under the influence of a fixed, out-of-plane field acting on the center of the stripe. (Color online). \label{fig:fixed_field}}
\end{figure}

Further confirmation of the non-trivial nature of the DW-vortex coupling is obtained from simulations where, instead of a disk in a vortex ground state, a static localized out-of-plane magnetic field mimicking the vortex core out-of-plane stray field is applied above the stripe. The hysteresis loops shown in Fig. \ref{fig:fixed_field} correspond to simulations where this static field has negative polarity (equivalent to a $p = -1$ vortex). They show asymmetric reversal, with strong depinning fields for positive applied field (antiparallel to the static field) while for negative applied fields the loops have no extraordinary features. By reversing the polarity of this static field, the depinning processes will appear on the negative side of the loop, with the positive side not showing any interesting features (data not shown). These polarity-dependent depinning fields, and the consequent asymmetries in the magnetization reversal, are reminiscent of situations where DWs are magnetostatically coupled to static pinning fields \cite{dots1, dots2}. However, the loops in Fig. \ref{fig:fixed_field} clearly show that it is not sufficient to consider only the coupling between the DW and the out-of-plane component of the VC stray field as source of DW pinning. When only this static out-of-plane field is present, the DW depinning fields are always weaker than the depinning fields of the DW coupled to the full vortex, showing that the in-plane magnetostatic coupling also plays a significant role in the overall DW pinning process, and that the evolution of the magnetization of the disk while interacting with the DW cannot be neglected.

\begin{figure}
\centering
\includegraphics[width=0.6\textwidth]{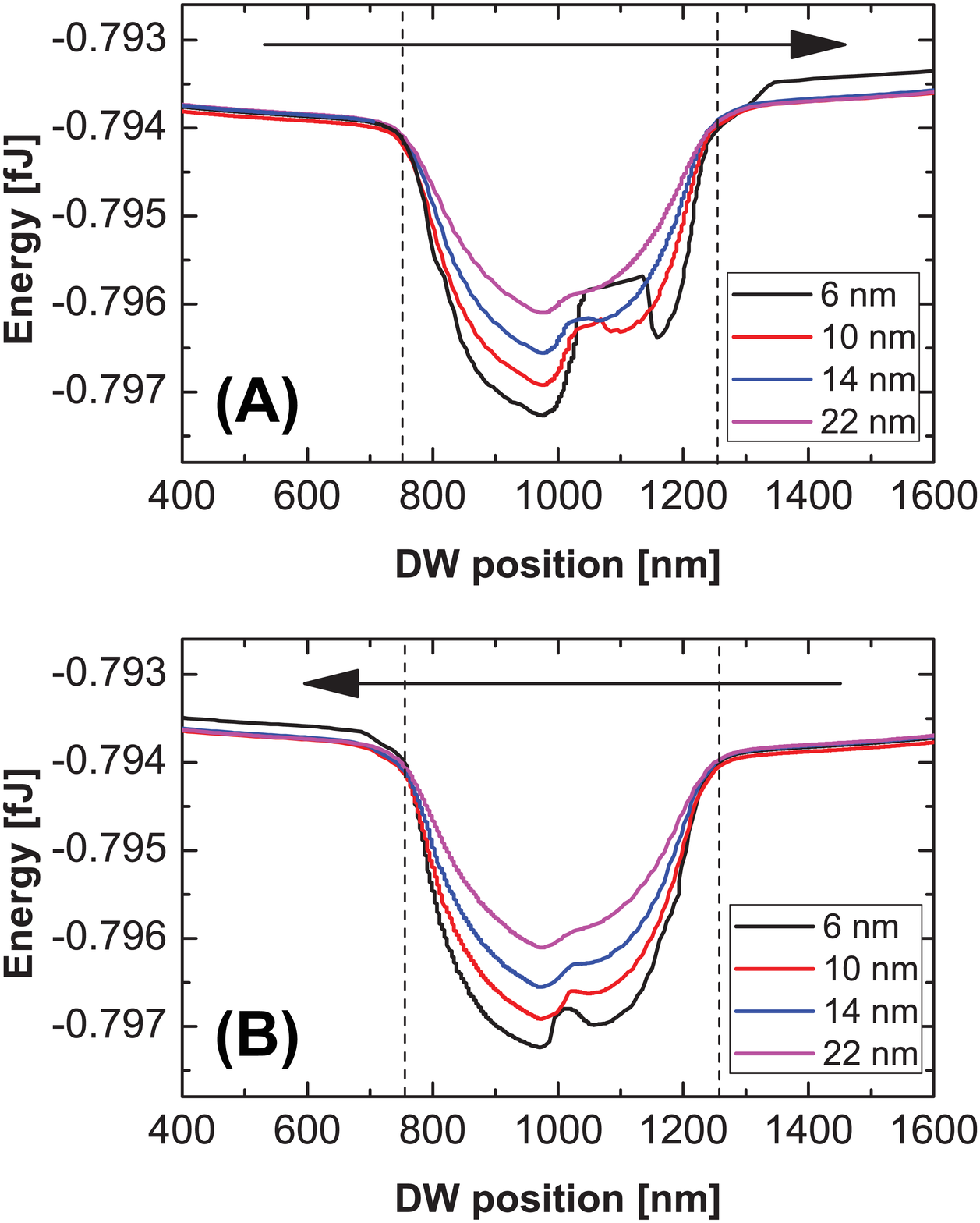}
\caption{Energy landscapes of a DW moving under the influence of a vortex with $c = +1$ and $p = -1$ from (a) left to right, with positive out-of-plane applied field, and (b) right to left, with negative out-of-plane applied field. The vertical dashed lines correspond to the disk edges. (Color online). \label{fig:energies}}
\end{figure}

These two contributions to the magnetostatic DW-vortex coupling behind the observed DW pinning may be better understood with the aid of the energy landscape of the system. In Fig. \ref{fig:energies}, the sum of the exchange, magnetostatic and anisotropy energies is plotted against the DW position along the stripe for $6$, $10$, $14$ and $22$ nm thick spacers and both positive and negative out-of-plane applied fields. As the DW approaches the edge of the disk (for $H_z > 0$) the energy decreases, forming a potential well which confirms the energetically favorable in-plane magnetostatic coupling between the DW and the vortex. In these regions close to the disk edges the potential well is symmetric, indicating that the in-plane coupling does not depend on the sign of the applied field (consequently, on the sense of DW motion). As the DW moves further left and reaches the core position (near the middle of the stripe), the energy increases, indicating that the VC effectively acts as an energy barrier for DW propagation. This energy increase is very sharp for the $6$ nm spacer, but gets weaker for thicker spacers, becoming barely visible when the spacer thickness is $22$ nm. Furthermore, when the spacer thickness increases, the symmetric potential well becomes less deep, consequence of the thickness-dependence of the magnetostatic coupling.

Under a negative applied field ($H_z < 0$), a DW located at the right edge of the stripe will move towards the left, and the symmetric potential well is still present. However, the VC-induced energy barrier is smaller in this case, making it easier for the DW to move past the VC position. This is the origin of the asymmetric reversal evidenced by the hysteresis loops shown in Figs. \ref{fig:cycles} and \ref{fig:thicknesses}. Notice that as the spacer layer thickness increases, not only the vortex-core energy barrier becomes less pronounced, but the energy landscape becomes nearly independent of the applied field polarity ($22$ nm curves in Fig. \ref{fig:energies}), leading to a more symmetric magnetization reversal process, as evidenced by the $22$ nm hysteresis loop in Fig. \ref{fig:thicknesses}. The analysis of the energy landscapes thus explains the main characteristics of the simulated hysteresis loops, namely: a symmetric broadening caused by the in-plane coupling, and an asymmetric reversal caused by out-of-plane coupling to the VC.

\begin{figure}
\centering
\includegraphics[width=0.75\textwidth]{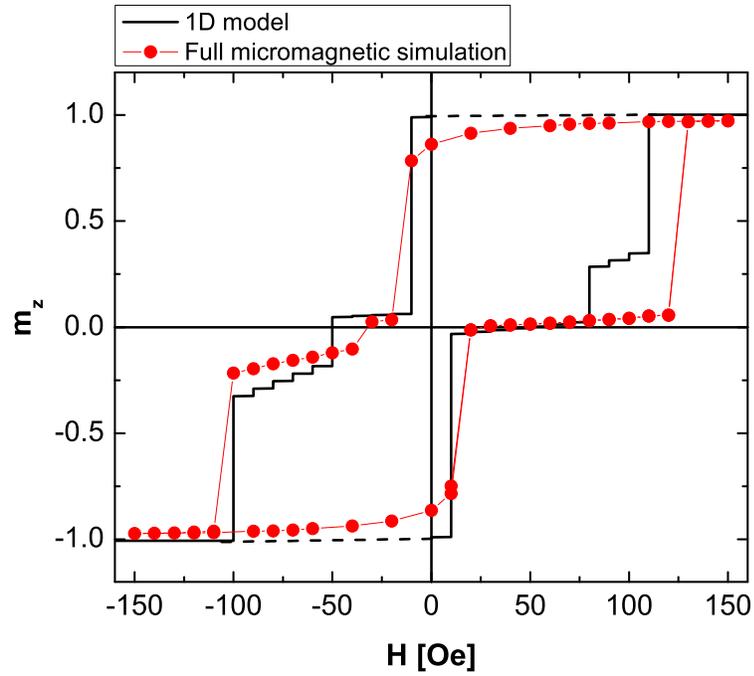}
\caption{Hysteresis loops for a 6 nm spacer obtained from the 1D model (line) and from a full micromagnetic simulation (red circles). Most features of the full 3D micromagnetic simulation are reproduced by a simplified 1D model, notably the two couplings (in-plane and out-of-plane) leading to DW pinning. (Color online). \label{fig:1D_model}}
\end{figure}

The energy landscapes in Fig. \ref{fig:energies} allowed us to develop a 1D model for the propagation of the DW along the stripe \cite{slonczewski, thomas, emori, lo_conte}. In this model, the DW dynamics is described in terms of DW position $q$ and the DW angle $\psi$ by the following equations:

\begin{eqnarray}
  (1 + \alpha^2) \frac{d q}{d t} = \alpha \gamma \Delta H_z + \frac{1}{2} \gamma \Delta H_k \sin{2 \psi} - \frac{\alpha \gamma \Delta}{2 M_s L_y L_z}\Big( \frac{d V}{d q}\Big) \\
  (1 + \alpha^2) \frac{d \psi}{d t} = \gamma H_z - \frac{1}{2} \alpha \gamma H_k \sin{2 \psi} - \frac{ \gamma}{2 M_s L_y L_z}\Big( \frac{d V}{d q}\Big)
\end{eqnarray}

where $\Delta = (A/K_{eff})^{1/2} = 6.3$ nm is the DW width, with $K_{eff} = K_u - 2 \pi M_s^2$, $K_u$ is the perpendicular uniaxial anisotropy constant, $M_s$ is the saturation magnetization, $A$ is the exchange stiffness constant, $\gamma$ is the gyromagnetic ratio, $H_k = N_x M_s$ is the shape anisotropy field with $N_x = L_z \ln(2)/(\pi \Delta)$ being the demagnetizing factor, $\alpha$ is the Gilbert damping parameter, $L_y$ and $L_z$ are the width and the thickness of the stripe and $V$ is the pinning potential from Fig. \ref{fig:energies}, approximated by a superposition of elementary functions. All these values were taken from the micromagnetic model defined in Sec. \ref{sims}. The model was used to simulate DW propagation for $6$, $14$ and $22$ nm spacer layers. The resulting hysteresis loop for a $6$ nm spacer is shown in Fig. \ref{fig:1D_model}, along with a hysteresis loop from a full 3D micromagnetic simulation. Despite the crudeness of this 1D model, which ignores the 3D character of the DW-vortex coupling unveiled by the micromagnetic simulations, the main features of the magnetization reversal of the stripe are reproduced, namely: the free propagation under a low field, the pinning under the dot edge, the higher fields necessary to propagate the DW under the disk and the pinning caused by the VC stray field.


\section{Conclusion \label{concl}}

It has been demonstrated that magnetic vortices in nanosized disks can effectively pin DWs moving along magnetic stripes with perpendicular anisotropy. Pinning can be achieved entirely by means of magnetostatic coupling between the two structures. The coupling is dependent on both the out-of-plane and the in-plane components of the stray fields and magnetizations of the DW and the vortex, giving rise to a complex pinning scenario which leads to simple phenomenology: asymmetric, broadened hysteresis loops of the magnetic stripes, with their magnetization reversal being easier when the out-of-plane applied field and the VC magnetization are parallel, and harder when they are antiparallel; and the broadening arising from the symmetric in-plane coupling between the DW and the vortex, which is independent of the DW sense of motion. The observed behavior cannot be mimicked by a DW simply coupled to a static out-of-plane field emulating the VC stray field, emphasizing the role of the in-plane coupling between DW stray fields and the vortex. A simplified 1D model of DW propagation where an approximation of the micromagnetic energy landscape, taking into account both in-plane and out-of-plane magnetostatic couplings, is able to qualitatively reproduce the main features of the magnetization process unveiled by the micromagnetic simulations. It is important to stress that in previous works \cite{dots1, dots2, franken, van_mourik}, the DW was in general coupled to nanostructures in a single domain state, from which trivial in-plane \cite{van_mourik} or out-of-plane \cite{franken} stray fields emanated. This lies in contrast to the present study, where a magnetic structure (vortex) from which no in-plane stray field emanates is used to influence the DW motion. These results may be useful to introduce a new way to achieve control of DW or vortex dynamics in spintronic devices. Furthermore, they can serve as a proof of principle and guide experimental efforts to fabricate similar structures and study their magnetization process.

\begin{acknowledgements}
RLN acknowledges the Brazilian agencies CNPq and FAPERJ for the postdoctoral grant during part of this work. The authors thank Dr. P. J. Metaxas for useful discussions, and Dr. A. Torres for granting access to the UFSC Physics Department computational facilities.
\end{acknowledgements}


\section*{References}
\bibliography{vortex_pinning_refs}

\begin{thebibliography}{29}%
\makeatletter
\providecommand \@ifxundefined [1]{%
 \@ifx{#1\undefined}
}%
\providecommand \@ifnum [1]{%
 \ifnum #1\expandafter \@firstoftwo
 \else \expandafter \@secondoftwo
 \fi
}%
\providecommand \@ifx [1]{%
 \ifx #1\expandafter \@firstoftwo
 \else \expandafter \@secondoftwo
 \fi
}%
\providecommand \natexlab [1]{#1}%
\providecommand \enquote  [1]{``#1''}%
\providecommand \bibnamefont  [1]{#1}%
\providecommand \bibfnamefont [1]{#1}%
\providecommand \citenamefont [1]{#1}%
\providecommand \href@noop [0]{\@secondoftwo}%
\providecommand \href [0]{\begingroup \@sanitize@url \@href}%
\providecommand \@href[1]{\@@startlink{#1}\@@href}%
\providecommand \@@href[1]{\endgroup#1\@@endlink}%
\providecommand \@sanitize@url [0]{\catcode `\\12\catcode `\$12\catcode
  `\&12\catcode `\#12\catcode `\^12\catcode `\_12\catcode `\%12\relax}%
\providecommand \@@startlink[1]{}%
\providecommand \@@endlink[0]{}%
\providecommand \url  [0]{\begingroup\@sanitize@url \@url }%
\providecommand \@url [1]{\endgroup\@href {#1}{\urlprefix }}%
\providecommand \urlprefix  [0]{URL }%
\providecommand \Eprint [0]{\href }%
\providecommand \doibase [0]{http://dx.doi.org/}%
\providecommand \selectlanguage [0]{\@gobble}%
\providecommand \bibinfo  [0]{\@secondoftwo}%
\providecommand \bibfield  [0]{\@secondoftwo}%
\providecommand \translation [1]{[#1]}%
\providecommand \BibitemOpen [0]{}%
\providecommand \bibitemStop [0]{}%
\providecommand \bibitemNoStop [0]{.\EOS\space}%
\providecommand \EOS [0]{\spacefactor3000\relax}%
\providecommand \BibitemShut  [1]{\csname bibitem#1\endcsname}%
\let\auto@bib@innerbib\@empty
\bibitem [{\citenamefont {Parkin}\ \emph {et~al.}(2008)\citenamefont {Parkin},
  \citenamefont {Hayashi},\ and\ \citenamefont {Thomas}}]{parkin}%
  \BibitemOpen
  \bibfield  {author} {\bibinfo {author} {\bibfnamefont {S.~S.~P.}\
  \bibnamefont {Parkin}}, \bibinfo {author} {\bibfnamefont {M.}~\bibnamefont
  {Hayashi}}, \ and\ \bibinfo {author} {\bibfnamefont {L.}~\bibnamefont
  {Thomas}},\ }\href@noop {} {\bibfield  {journal} {\bibinfo  {journal}
  {Science}\ }\textbf {\bibinfo {volume} {320}},\ \bibinfo {pages} {190}
  (\bibinfo {year} {2008})}\BibitemShut {NoStop}%
\bibitem [{\citenamefont {Allwood}\ \emph {et~al.}(2005)\citenamefont
  {Allwood}, \citenamefont {Xiong}, \citenamefont {Faulkner}, \citenamefont
  {Atkinson}, \citenamefont {Petit},\ and\ \citenamefont
  {Cowburn}}]{allwood_logic}%
  \BibitemOpen
  \bibfield  {author} {\bibinfo {author} {\bibfnamefont {D.}~\bibnamefont
  {Allwood}}, \bibinfo {author} {\bibfnamefont {G.}~\bibnamefont {Xiong}},
  \bibinfo {author} {\bibfnamefont {C.}~\bibnamefont {Faulkner}}, \bibinfo
  {author} {\bibfnamefont {D.}~\bibnamefont {Atkinson}}, \bibinfo {author}
  {\bibfnamefont {D.}~\bibnamefont {Petit}}, \ and\ \bibinfo {author}
  {\bibfnamefont {R.}~\bibnamefont {Cowburn}},\ }\href@noop {} {\bibfield
  {journal} {\bibinfo  {journal} {Science}\ }\textbf {\bibinfo {volume}
  {309}},\ \bibinfo {pages} {1688} (\bibinfo {year} {2005})}\BibitemShut
  {NoStop}%
\bibitem [{\citenamefont {Hrkac}\ \emph {et~al.}(2011)\citenamefont {Hrkac},
  \citenamefont {Dean},\ and\ \citenamefont {Allwood}}]{allwood_review}%
  \BibitemOpen
  \bibfield  {author} {\bibinfo {author} {\bibfnamefont {G.}~\bibnamefont
  {Hrkac}}, \bibinfo {author} {\bibfnamefont {J.}~\bibnamefont {Dean}}, \ and\
  \bibinfo {author} {\bibfnamefont {D.~A.}\ \bibnamefont {Allwood}},\
  }\href@noop {} {\bibfield  {journal} {\bibinfo  {journal} {Phil. Trans. R.
  Soc. A}\ }\textbf {\bibinfo {volume} {369}},\ \bibinfo {pages} {3214}
  (\bibinfo {year} {2011})}\BibitemShut {NoStop}%
\bibitem [{\citenamefont {Vieu}\ \emph {et~al.}(2002)\citenamefont {Vieu},
  \citenamefont {Gierak}, \citenamefont {Launois}, \citenamefont {Aign},
  \citenamefont {Meyer}, \citenamefont {Jamet}, \citenamefont {Ferr{\'e}},
  \citenamefont {Chappert}, \citenamefont {Devolder}, \citenamefont {Mathet},\
  and\ \citenamefont {Bernas}}]{vieu}%
  \BibitemOpen
  \bibfield  {author} {\bibinfo {author} {\bibfnamefont {C.}~\bibnamefont
  {Vieu}}, \bibinfo {author} {\bibfnamefont {J.}~\bibnamefont {Gierak}},
  \bibinfo {author} {\bibfnamefont {H.}~\bibnamefont {Launois}}, \bibinfo
  {author} {\bibfnamefont {T.}~\bibnamefont {Aign}}, \bibinfo {author}
  {\bibfnamefont {P.}~\bibnamefont {Meyer}}, \bibinfo {author} {\bibfnamefont
  {J.-P.}\ \bibnamefont {Jamet}}, \bibinfo {author} {\bibfnamefont
  {J.}~\bibnamefont {Ferr{\'e}}}, \bibinfo {author} {\bibfnamefont
  {C.}~\bibnamefont {Chappert}}, \bibinfo {author} {\bibfnamefont
  {T.}~\bibnamefont {Devolder}}, \bibinfo {author} {\bibfnamefont
  {V.}~\bibnamefont {Mathet}}, \ and\ \bibinfo {author} {\bibfnamefont
  {H.}~\bibnamefont {Bernas}},\ }\href@noop {} {\bibfield  {journal} {\bibinfo
  {journal} {J. Appl. Phys.}\ }\textbf {\bibinfo {volume} {91}},\ \bibinfo
  {pages} {3103} (\bibinfo {year} {2002})}\BibitemShut {NoStop}%
\bibitem [{\citenamefont {Repain}\ \emph {et~al.}(2004)\citenamefont {Repain},
  \citenamefont {Jamet}, \citenamefont {Vernier}, \citenamefont {Bauer},
  \citenamefont {Ferre}, \citenamefont {Chappert}, \citenamefont {Gierak},\
  and\ \citenamefont {Mailly}}]{repain}%
  \BibitemOpen
  \bibfield  {author} {\bibinfo {author} {\bibfnamefont {V.}~\bibnamefont
  {Repain}}, \bibinfo {author} {\bibfnamefont {J.}~\bibnamefont {Jamet}},
  \bibinfo {author} {\bibfnamefont {N.}~\bibnamefont {Vernier}}, \bibinfo
  {author} {\bibfnamefont {M.}~\bibnamefont {Bauer}}, \bibinfo {author}
  {\bibfnamefont {J.}~\bibnamefont {Ferre}}, \bibinfo {author} {\bibfnamefont
  {C.}~\bibnamefont {Chappert}}, \bibinfo {author} {\bibfnamefont
  {J.}~\bibnamefont {Gierak}}, \ and\ \bibinfo {author} {\bibfnamefont
  {D.}~\bibnamefont {Mailly}},\ }\href@noop {} {\bibfield  {journal} {\bibinfo
  {journal} {J. Appl. Phys.}\ }\textbf {\bibinfo {volume} {95}},\ \bibinfo
  {pages} {2614} (\bibinfo {year} {2004})}\BibitemShut {NoStop}%
\bibitem [{\citenamefont {Franken}\ \emph {et~al.}(2011)\citenamefont
  {Franken}, \citenamefont {Hoeijmakers}, \citenamefont {Lavrijsen},\ and\
  \citenamefont {Swagten}}]{franken_anisotropy}%
  \BibitemOpen
  \bibfield  {author} {\bibinfo {author} {\bibfnamefont {J.~H.}\ \bibnamefont
  {Franken}}, \bibinfo {author} {\bibfnamefont {M.}~\bibnamefont
  {Hoeijmakers}}, \bibinfo {author} {\bibfnamefont {R.}~\bibnamefont
  {Lavrijsen}}, \ and\ \bibinfo {author} {\bibfnamefont {H.~J.~M.}\
  \bibnamefont {Swagten}},\ }\href@noop {} {\bibfield  {journal} {\bibinfo
  {journal} {J. Phys.: Condens. Matter}\ }\textbf {\bibinfo {volume} {24}},\
  \bibinfo {pages} {024216} (\bibinfo {year} {2011})}\BibitemShut {NoStop}%
\bibitem [{\citenamefont {Adeyeye}\ \emph {et~al.}(1997)\citenamefont
  {Adeyeye}, \citenamefont {Bland},\ and\ \citenamefont {DABOO}}]{adeyeye}%
  \BibitemOpen
  \bibfield  {author} {\bibinfo {author} {\bibfnamefont {A.~O.}\ \bibnamefont
  {Adeyeye}}, \bibinfo {author} {\bibfnamefont {J.~A.~C.}\ \bibnamefont
  {Bland}}, \ and\ \bibinfo {author} {\bibfnamefont {C.}~\bibnamefont
  {DABOO}},\ }\href@noop {} {\bibfield  {journal} {\bibinfo  {journal} {Appl.
  Phys. Lett.}\ }\textbf {\bibinfo {volume} {70}},\ \bibinfo {pages} {3164}
  (\bibinfo {year} {1997})}\BibitemShut {NoStop}%
\bibitem [{\citenamefont {Lewis}\ \emph {et~al.}(2010)\citenamefont {Lewis},
  \citenamefont {Petit}, \citenamefont {O'Brien}, \citenamefont
  {Fernandez-Pacheco}, \citenamefont {Sampaio}, \citenamefont {Jausovec},
  \citenamefont {Zeng}, \citenamefont {Read},\ and\ \citenamefont
  {Cowburn}}]{cowburn_comb}%
  \BibitemOpen
  \bibfield  {author} {\bibinfo {author} {\bibfnamefont {E.~R.}\ \bibnamefont
  {Lewis}}, \bibinfo {author} {\bibfnamefont {D.}~\bibnamefont {Petit}},
  \bibinfo {author} {\bibfnamefont {L.}~\bibnamefont {O'Brien}}, \bibinfo
  {author} {\bibfnamefont {A.}~\bibnamefont {Fernandez-Pacheco}}, \bibinfo
  {author} {\bibfnamefont {J.}~\bibnamefont {Sampaio}}, \bibinfo {author}
  {\bibfnamefont {A.-V.}\ \bibnamefont {Jausovec}}, \bibinfo {author}
  {\bibfnamefont {H.~T.}\ \bibnamefont {Zeng}}, \bibinfo {author}
  {\bibfnamefont {D.~E.}\ \bibnamefont {Read}}, \ and\ \bibinfo {author}
  {\bibfnamefont {R.~P.}\ \bibnamefont {Cowburn}},\ }\href@noop {} {\bibfield
  {journal} {\bibinfo  {journal} {Nat. Mater.}\ }\textbf {\bibinfo {volume}
  {9}},\ \bibinfo {pages} {980} (\bibinfo {year} {2010})}\BibitemShut {NoStop}%
\bibitem [{\citenamefont {Hayashi}\ \emph {et~al.}(2006)\citenamefont
  {Hayashi}, \citenamefont {Thomas}, \citenamefont {Rettner}, \citenamefont
  {Moriya}, \citenamefont {Jiang},\ and\ \citenamefont
  {Parkin}}]{hayashi_notch}%
  \BibitemOpen
  \bibfield  {author} {\bibinfo {author} {\bibfnamefont {M.}~\bibnamefont
  {Hayashi}}, \bibinfo {author} {\bibfnamefont {L.}~\bibnamefont {Thomas}},
  \bibinfo {author} {\bibfnamefont {C.}~\bibnamefont {Rettner}}, \bibinfo
  {author} {\bibfnamefont {R.}~\bibnamefont {Moriya}}, \bibinfo {author}
  {\bibfnamefont {X.}~\bibnamefont {Jiang}}, \ and\ \bibinfo {author}
  {\bibfnamefont {S.~S.~P.}\ \bibnamefont {Parkin}},\ }\href@noop {} {\bibfield
   {journal} {\bibinfo  {journal} {Phys. Rev. Lett.}\ }\textbf {\bibinfo
  {volume} {97}},\ \bibinfo {pages} {207205} (\bibinfo {year}
  {2006})}\BibitemShut {NoStop}%
\bibitem [{\citenamefont {Hayashi}\ \emph {et~al.}(2007)\citenamefont
  {Hayashi}, \citenamefont {Thomas}, \citenamefont {Rettner}, \citenamefont
  {Moriya},\ and\ \citenamefont {Parkin}}]{hayashi_notch2}%
  \BibitemOpen
  \bibfield  {author} {\bibinfo {author} {\bibfnamefont {M.}~\bibnamefont
  {Hayashi}}, \bibinfo {author} {\bibfnamefont {L.}~\bibnamefont {Thomas}},
  \bibinfo {author} {\bibfnamefont {C.}~\bibnamefont {Rettner}}, \bibinfo
  {author} {\bibfnamefont {R.}~\bibnamefont {Moriya}}, \ and\ \bibinfo {author}
  {\bibfnamefont {S.~S.~P.}\ \bibnamefont {Parkin}},\ }\href@noop {} {\bibfield
   {journal} {\bibinfo  {journal} {Nat. Phys.}\ }\textbf {\bibinfo {volume}
  {3}},\ \bibinfo {pages} {21} (\bibinfo {year} {2007})}\BibitemShut {NoStop}%
\bibitem [{\citenamefont {Bogart}\ \emph {et~al.}(2009)\citenamefont {Bogart},
  \citenamefont {Atkinson}, \citenamefont {O'Shea}, \citenamefont
  {McGrouther},\ and\ \citenamefont {McVitie}}]{atkinson_notch}%
  \BibitemOpen
  \bibfield  {author} {\bibinfo {author} {\bibfnamefont {L.~K.}\ \bibnamefont
  {Bogart}}, \bibinfo {author} {\bibfnamefont {D.}~\bibnamefont {Atkinson}},
  \bibinfo {author} {\bibfnamefont {K.}~\bibnamefont {O'Shea}}, \bibinfo
  {author} {\bibfnamefont {D.}~\bibnamefont {McGrouther}}, \ and\ \bibinfo
  {author} {\bibfnamefont {S.}~\bibnamefont {McVitie}},\ }\href@noop {}
  {\bibfield  {journal} {\bibinfo  {journal} {Phys. Rev. B}\ }\textbf {\bibinfo
  {volume} {79}},\ \bibinfo {pages} {054414} (\bibinfo {year}
  {2009})}\BibitemShut {NoStop}%
\bibitem [{\citenamefont {Brand{\~a}o}\ \emph {et~al.}(2014)\citenamefont
  {Brand{\~a}o}, \citenamefont {Novak}, \citenamefont {Lozano}, \citenamefont
  {Soledade}, \citenamefont {Mello}, \citenamefont {Garcia},\ and\
  \citenamefont {Sampaio}}]{brandao_novak}%
  \BibitemOpen
  \bibfield  {author} {\bibinfo {author} {\bibfnamefont {J.}~\bibnamefont
  {Brand{\~a}o}}, \bibinfo {author} {\bibfnamefont {R.~L.}\ \bibnamefont
  {Novak}}, \bibinfo {author} {\bibfnamefont {H.}~\bibnamefont {Lozano}},
  \bibinfo {author} {\bibfnamefont {P.~R.}\ \bibnamefont {Soledade}}, \bibinfo
  {author} {\bibfnamefont {A.}~\bibnamefont {Mello}}, \bibinfo {author}
  {\bibfnamefont {F.}~\bibnamefont {Garcia}}, \ and\ \bibinfo {author}
  {\bibfnamefont {L.~C.}\ \bibnamefont {Sampaio}},\ }\href@noop {} {\bibfield
  {journal} {\bibinfo  {journal} {J. Appl. Phys.}\ }\textbf {\bibinfo {volume}
  {116}},\ \bibinfo {pages} {193902} (\bibinfo {year} {2014})}\BibitemShut
  {NoStop}%
\bibitem [{\citenamefont {Dean}\ \emph {et~al.}(2015)\citenamefont {Dean},
  \citenamefont {Bryan}, \citenamefont {Cooper}, \citenamefont {Virbule},
  \citenamefont {Cunningham},\ and\ \citenamefont {Hayward}}]{sound_idea}%
  \BibitemOpen
  \bibfield  {author} {\bibinfo {author} {\bibfnamefont {J.}~\bibnamefont
  {Dean}}, \bibinfo {author} {\bibfnamefont {M.~T.}\ \bibnamefont {Bryan}},
  \bibinfo {author} {\bibfnamefont {J.~D.}\ \bibnamefont {Cooper}}, \bibinfo
  {author} {\bibfnamefont {A.}~\bibnamefont {Virbule}}, \bibinfo {author}
  {\bibfnamefont {J.~E.}\ \bibnamefont {Cunningham}}, \ and\ \bibinfo {author}
  {\bibfnamefont {T.~J.}\ \bibnamefont {Hayward}},\ }\href@noop {} {\bibfield
  {journal} {\bibinfo  {journal} {Appl. Phys. Lett.}\ }\textbf {\bibinfo
  {volume} {107}},\ \bibinfo {pages} {142405} (\bibinfo {year}
  {2015})}\BibitemShut {NoStop}%
\bibitem [{\citenamefont {Metaxas}\ \emph {et~al.}(2013)\citenamefont
  {Metaxas}, \citenamefont {Zermatten}, \citenamefont {Novak}, \citenamefont
  {Rohart}, \citenamefont {Jamet}, \citenamefont {Weil}, \citenamefont
  {Ferr{\'e}}, \citenamefont {Mougin}, \citenamefont {Stamps}, \citenamefont
  {Gaudin}, \citenamefont {Baltz},\ and\ \citenamefont {Rodmacq}}]{dots1}%
  \BibitemOpen
  \bibfield  {author} {\bibinfo {author} {\bibfnamefont {P.~J.}\ \bibnamefont
  {Metaxas}}, \bibinfo {author} {\bibfnamefont {P.-J.}\ \bibnamefont
  {Zermatten}}, \bibinfo {author} {\bibfnamefont {R.~L.}\ \bibnamefont
  {Novak}}, \bibinfo {author} {\bibfnamefont {S.}~\bibnamefont {Rohart}},
  \bibinfo {author} {\bibfnamefont {J.-P.}\ \bibnamefont {Jamet}}, \bibinfo
  {author} {\bibfnamefont {R.}~\bibnamefont {Weil}}, \bibinfo {author}
  {\bibfnamefont {J.}~\bibnamefont {Ferr{\'e}}}, \bibinfo {author}
  {\bibfnamefont {A.}~\bibnamefont {Mougin}}, \bibinfo {author} {\bibfnamefont
  {R.~L.}\ \bibnamefont {Stamps}}, \bibinfo {author} {\bibfnamefont
  {G.}~\bibnamefont {Gaudin}}, \bibinfo {author} {\bibfnamefont
  {V.}~\bibnamefont {Baltz}}, \ and\ \bibinfo {author} {\bibfnamefont
  {B.}~\bibnamefont {Rodmacq}},\ }\href@noop {} {\bibfield  {journal} {\bibinfo
   {journal} {J. Appl. Phys.}\ }\textbf {\bibinfo {volume} {113}},\ \bibinfo
  {pages} {073906} (\bibinfo {year} {2013})}\BibitemShut {NoStop}%
\bibitem [{\citenamefont {Novak}\ \emph {et~al.}(2015)\citenamefont {Novak},
  \citenamefont {Metaxas}, \citenamefont {Jamet}, \citenamefont {Weil},
  \citenamefont {Ferr{\'e}}, \citenamefont {Mougin}, \citenamefont {Rohart},
  \citenamefont {Stamps}, \citenamefont {Zermatten}, \citenamefont {Gaudin},
  \citenamefont {Baltz},\ and\ \citenamefont {Rodmacq}}]{dots2}%
  \BibitemOpen
  \bibfield  {author} {\bibinfo {author} {\bibfnamefont {R.~L.}\ \bibnamefont
  {Novak}}, \bibinfo {author} {\bibfnamefont {P.~J.}\ \bibnamefont {Metaxas}},
  \bibinfo {author} {\bibfnamefont {J.-P.}\ \bibnamefont {Jamet}}, \bibinfo
  {author} {\bibfnamefont {R.}~\bibnamefont {Weil}}, \bibinfo {author}
  {\bibfnamefont {J.}~\bibnamefont {Ferr{\'e}}}, \bibinfo {author}
  {\bibfnamefont {A.}~\bibnamefont {Mougin}}, \bibinfo {author} {\bibfnamefont
  {S.}~\bibnamefont {Rohart}}, \bibinfo {author} {\bibfnamefont {R.~L.}\
  \bibnamefont {Stamps}}, \bibinfo {author} {\bibfnamefont {P.-J.}\
  \bibnamefont {Zermatten}}, \bibinfo {author} {\bibfnamefont {G.}~\bibnamefont
  {Gaudin}}, \bibinfo {author} {\bibfnamefont {V.}~\bibnamefont {Baltz}}, \
  and\ \bibinfo {author} {\bibfnamefont {B.}~\bibnamefont {Rodmacq}},\
  }\href@noop {} {\bibfield  {journal} {\bibinfo  {journal} {J. Phys. D: Appl.
  Phys.}\ }\textbf {\bibinfo {volume} {48}},\ \bibinfo {pages} {235004}
  (\bibinfo {year} {2015})}\BibitemShut {NoStop}%
\bibitem [{\citenamefont {Franken}\ \emph {et~al.}(2014)\citenamefont
  {Franken}, \citenamefont {van~der Heijden}, \citenamefont {Ellis},
  \citenamefont {Lavrijsen}, \citenamefont {Daniels}, \citenamefont
  {McGrouther}, \citenamefont {Swagten},\ and\ \citenamefont
  {Koopmans}}]{franken}%
  \BibitemOpen
  \bibfield  {author} {\bibinfo {author} {\bibfnamefont {J.~H.}\ \bibnamefont
  {Franken}}, \bibinfo {author} {\bibfnamefont {M.~A.~J.}\ \bibnamefont
  {van~der Heijden}}, \bibinfo {author} {\bibfnamefont {T.~H.}\ \bibnamefont
  {Ellis}}, \bibinfo {author} {\bibfnamefont {R.}~\bibnamefont {Lavrijsen}},
  \bibinfo {author} {\bibfnamefont {C.}~\bibnamefont {Daniels}}, \bibinfo
  {author} {\bibfnamefont {D.}~\bibnamefont {McGrouther}}, \bibinfo {author}
  {\bibfnamefont {H.~J.~M.}\ \bibnamefont {Swagten}}, \ and\ \bibinfo {author}
  {\bibfnamefont {B.}~\bibnamefont {Koopmans}},\ }\href@noop {} {\bibfield
  {journal} {\bibinfo  {journal} {Adv. Funct. Mater.}\ }\textbf {\bibinfo
  {volume} {24}},\ \bibinfo {pages} {3508} (\bibinfo {year}
  {2014})}\BibitemShut {NoStop}%
\bibitem [{\citenamefont {van Mourik}\ \emph {et~al.}(2014)\citenamefont {van
  Mourik}, \citenamefont {Rettner}, \citenamefont {Koopmans},\ and\
  \citenamefont {Parkin}}]{van_mourik}%
  \BibitemOpen
  \bibfield  {author} {\bibinfo {author} {\bibfnamefont {R.~A.}\ \bibnamefont
  {van Mourik}}, \bibinfo {author} {\bibfnamefont {C.~T.}\ \bibnamefont
  {Rettner}}, \bibinfo {author} {\bibfnamefont {B.}~\bibnamefont {Koopmans}}, \
  and\ \bibinfo {author} {\bibfnamefont {S.~S.~P.}\ \bibnamefont {Parkin}},\
  }\href@noop {} {\bibfield  {journal} {\bibinfo  {journal} {J. Appl. Phys.}\
  }\textbf {\bibinfo {volume} {115}},\ \bibinfo {pages} {17D503} (\bibinfo
  {year} {2014})}\BibitemShut {NoStop}%
\bibitem [{\citenamefont {Cowburn}\ \emph {et~al.}(1999)\citenamefont
  {Cowburn}, \citenamefont {Koltsov}, \citenamefont {Adeyeye}, \citenamefont
  {Welland},\ and\ \citenamefont {Tricker}}]{cowburn}%
  \BibitemOpen
  \bibfield  {author} {\bibinfo {author} {\bibfnamefont {R.}~\bibnamefont
  {Cowburn}}, \bibinfo {author} {\bibfnamefont {D.}~\bibnamefont {Koltsov}},
  \bibinfo {author} {\bibfnamefont {A.}~\bibnamefont {Adeyeye}}, \bibinfo
  {author} {\bibfnamefont {M.}~\bibnamefont {Welland}}, \ and\ \bibinfo
  {author} {\bibfnamefont {D.}~\bibnamefont {Tricker}},\ }\href@noop {}
  {\bibfield  {journal} {\bibinfo  {journal} {Phys. Rev. Lett.}\ }\textbf
  {\bibinfo {volume} {83}},\ \bibinfo {pages} {1042} (\bibinfo {year}
  {1999})}\BibitemShut {NoStop}%
\bibitem [{\citenamefont {Shinjo}\ \emph {et~al.}(2000)\citenamefont {Shinjo},
  \citenamefont {Okuno}, \citenamefont {Hassdorf}, \citenamefont {Shigeto},\
  and\ \citenamefont {Ono}}]{shinjo}%
  \BibitemOpen
  \bibfield  {author} {\bibinfo {author} {\bibfnamefont {T.}~\bibnamefont
  {Shinjo}}, \bibinfo {author} {\bibfnamefont {T.}~\bibnamefont {Okuno}},
  \bibinfo {author} {\bibfnamefont {R.}~\bibnamefont {Hassdorf}}, \bibinfo
  {author} {\bibfnamefont {K.}~\bibnamefont {Shigeto}}, \ and\ \bibinfo
  {author} {\bibfnamefont {T.}~\bibnamefont {Ono}},\ }\href@noop {} {\bibfield
  {journal} {\bibinfo  {journal} {Science}\ }\textbf {\bibinfo {volume}
  {289}},\ \bibinfo {pages} {930} (\bibinfo {year} {2000})}\BibitemShut
  {NoStop}%
\bibitem [{\citenamefont {Wachowiak}\ \emph {et~al.}(2002)\citenamefont
  {Wachowiak}, \citenamefont {Wiebe}, \citenamefont {Bode}, \citenamefont
  {Pietzsch}, \citenamefont {Morgenstern},\ and\ \citenamefont
  {Wiesendanger}}]{wachowiak}%
  \BibitemOpen
  \bibfield  {author} {\bibinfo {author} {\bibfnamefont {A.}~\bibnamefont
  {Wachowiak}}, \bibinfo {author} {\bibfnamefont {J.}~\bibnamefont {Wiebe}},
  \bibinfo {author} {\bibfnamefont {M.}~\bibnamefont {Bode}}, \bibinfo {author}
  {\bibfnamefont {O.}~\bibnamefont {Pietzsch}}, \bibinfo {author}
  {\bibfnamefont {M.}~\bibnamefont {Morgenstern}}, \ and\ \bibinfo {author}
  {\bibfnamefont {R.}~\bibnamefont {Wiesendanger}},\ }\href@noop {} {\bibfield
  {journal} {\bibinfo  {journal} {Science}\ }\textbf {\bibinfo {volume}
  {298}},\ \bibinfo {pages} {577} (\bibinfo {year} {2002})}\BibitemShut
  {NoStop}%
\bibitem [{\citenamefont {{Alex Hubert and Rudolf
  Sch\"afer}}(1998)}]{hubert_book}%
  \BibitemOpen
  \bibfield  {author} {\bibinfo {author} {\bibnamefont {{Alex Hubert and Rudolf
  Sch\"afer}}},\ }\href@noop {} {\emph {\bibinfo {title} {Magnetic Domains: The
  Analysis of Magnetic Microstructures}}}\ (\bibinfo  {publisher} {Springer},\
  \bibinfo {year} {1998})\BibitemShut {NoStop}%
\bibitem [{\citenamefont {Feldtkeller}\ and\ \citenamefont
  {Thomas}(1965)}]{feldtkeller}%
  \BibitemOpen
  \bibfield  {author} {\bibinfo {author} {\bibfnamefont {E.}~\bibnamefont
  {Feldtkeller}}\ and\ \bibinfo {author} {\bibfnamefont {H.}~\bibnamefont
  {Thomas}},\ }\href@noop {} {\bibfield  {journal} {\bibinfo  {journal} {Phys.
  kondens. Materie}\ }\textbf {\bibinfo {volume} {4}},\ \bibinfo {pages} {8}
  (\bibinfo {year} {1965})}\BibitemShut {NoStop}%
\bibitem [{\citenamefont {Heldt}\ \emph {et~al.}(2014)\citenamefont {Heldt},
  \citenamefont {Bryan}, \citenamefont {Hrkac}, \citenamefont {Stevenson},
  \citenamefont {Chopdekar}, \citenamefont {Raabe}, \citenamefont {Thomson},\
  and\ \citenamefont {Heyderman}}]{heldt}%
  \BibitemOpen
  \bibfield  {author} {\bibinfo {author} {\bibfnamefont {G.}~\bibnamefont
  {Heldt}}, \bibinfo {author} {\bibfnamefont {M.~T.}\ \bibnamefont {Bryan}},
  \bibinfo {author} {\bibfnamefont {G.}~\bibnamefont {Hrkac}}, \bibinfo
  {author} {\bibfnamefont {S.~E.}\ \bibnamefont {Stevenson}}, \bibinfo {author}
  {\bibfnamefont {R.~V.}\ \bibnamefont {Chopdekar}}, \bibinfo {author}
  {\bibfnamefont {J.}~\bibnamefont {Raabe}}, \bibinfo {author} {\bibfnamefont
  {T.}~\bibnamefont {Thomson}}, \ and\ \bibinfo {author} {\bibfnamefont
  {L.~J.}\ \bibnamefont {Heyderman}},\ }\href@noop {} {\bibfield  {journal}
  {\bibinfo  {journal} {Appl. Phys. Lett.}\ }\textbf {\bibinfo {volume}
  {104}},\ \bibinfo {pages} {182401} (\bibinfo {year} {2014})}\BibitemShut
  {NoStop}%
\bibitem [{\citenamefont {Wohlh\"uter}\ \emph {et~al.}(2015)\citenamefont
  {Wohlh\"uter}, \citenamefont {Bryan}, \citenamefont {Warnicke}, \citenamefont
  {Gliga}, \citenamefont {Stevenson}, \citenamefont {Heldt}, \citenamefont
  {Saharan}, \citenamefont {Suszka}, \citenamefont {Moutafis}, \citenamefont
  {Chopdekar}, \citenamefont {Raabe}, \citenamefont {Thomson}, \citenamefont
  {Hrkac},\ and\ \citenamefont {Heyderman}}]{wohlhuter}%
  \BibitemOpen
  \bibfield  {author} {\bibinfo {author} {\bibfnamefont {P.}~\bibnamefont
  {Wohlh\"uter}}, \bibinfo {author} {\bibfnamefont {M.~T.}\ \bibnamefont
  {Bryan}}, \bibinfo {author} {\bibfnamefont {P.}~\bibnamefont {Warnicke}},
  \bibinfo {author} {\bibfnamefont {S.}~\bibnamefont {Gliga}}, \bibinfo
  {author} {\bibfnamefont {S.~E.}\ \bibnamefont {Stevenson}}, \bibinfo {author}
  {\bibfnamefont {G.}~\bibnamefont {Heldt}}, \bibinfo {author} {\bibfnamefont
  {L.}~\bibnamefont {Saharan}}, \bibinfo {author} {\bibfnamefont {A.~K.}\
  \bibnamefont {Suszka}}, \bibinfo {author} {\bibfnamefont {C.}~\bibnamefont
  {Moutafis}}, \bibinfo {author} {\bibfnamefont {R.~V.}\ \bibnamefont
  {Chopdekar}}, \bibinfo {author} {\bibfnamefont {J.}~\bibnamefont {Raabe}},
  \bibinfo {author} {\bibfnamefont {T.}~\bibnamefont {Thomson}}, \bibinfo
  {author} {\bibfnamefont {G.}~\bibnamefont {Hrkac}}, \ and\ \bibinfo {author}
  {\bibfnamefont {L.~J.}\ \bibnamefont {Heyderman}},\ }\href@noop {} {\bibfield
   {journal} {\bibinfo  {journal} {Nat. Comm.}\ }\textbf {\bibinfo {volume}
  {6}},\ \bibinfo {pages} {7836} (\bibinfo {year} {2015})}\BibitemShut
  {NoStop}%
\bibitem [{\citenamefont {Vansteenkiste}\ \emph {et~al.}(2014)\citenamefont
  {Vansteenkiste}, \citenamefont {Leliaert}, \citenamefont {Dvornik},
  \citenamefont {Helsen}, \citenamefont {Garcia-Sanchez},\ and\ \citenamefont
  {Van~Waeyenberge}}]{mumax3}%
  \BibitemOpen
  \bibfield  {author} {\bibinfo {author} {\bibfnamefont {A.}~\bibnamefont
  {Vansteenkiste}}, \bibinfo {author} {\bibfnamefont {J.}~\bibnamefont
  {Leliaert}}, \bibinfo {author} {\bibfnamefont {M.}~\bibnamefont {Dvornik}},
  \bibinfo {author} {\bibfnamefont {M.}~\bibnamefont {Helsen}}, \bibinfo
  {author} {\bibfnamefont {F.}~\bibnamefont {Garcia-Sanchez}}, \ and\ \bibinfo
  {author} {\bibfnamefont {B.}~\bibnamefont {Van~Waeyenberge}},\ }\href@noop {}
  {\bibfield  {journal} {\bibinfo  {journal} {AIP Advances}\ }\textbf {\bibinfo
  {volume} {4}},\ \bibinfo {pages} {107133} (\bibinfo {year}
  {2014})}\BibitemShut {NoStop}%
\bibitem [{\citenamefont {{Slonczewski, J. C. and Malozemoff, A.
  P.}}(1979)}]{slonczewski}%
  \BibitemOpen
  \bibfield  {author} {\bibinfo {author} {\bibnamefont {{Slonczewski, J. C. and
  Malozemoff, A. P.}}},\ }\href@noop {} {\emph {\bibinfo {title} {Magnetic
  Domain Walls in Bubble Materials}}}\ (\bibinfo  {publisher} {Academic
  Press},\ \bibinfo {year} {1979})\BibitemShut {NoStop}%
\bibitem [{\citenamefont {Thomas}\ \emph {et~al.}(2006)\citenamefont {Thomas},
  \citenamefont {Hayashi}, \citenamefont {Jiang}, \citenamefont {Moriya},
  \citenamefont {Rettner},\ and\ \citenamefont {Parkin}}]{thomas}%
  \BibitemOpen
  \bibfield  {author} {\bibinfo {author} {\bibfnamefont {L.}~\bibnamefont
  {Thomas}}, \bibinfo {author} {\bibfnamefont {M.}~\bibnamefont {Hayashi}},
  \bibinfo {author} {\bibfnamefont {X.}~\bibnamefont {Jiang}}, \bibinfo
  {author} {\bibfnamefont {R.}~\bibnamefont {Moriya}}, \bibinfo {author}
  {\bibfnamefont {C.}~\bibnamefont {Rettner}}, \ and\ \bibinfo {author}
  {\bibfnamefont {S.}~\bibnamefont {Parkin}},\ }\href@noop {} {\bibfield
  {journal} {\bibinfo  {journal} {Nature}\ }\textbf {\bibinfo {volume} {443}},\
  \bibinfo {pages} {197} (\bibinfo {year} {2006})}\BibitemShut {NoStop}%
\bibitem [{\citenamefont {Emori}(2013)}]{emori}%
  \BibitemOpen
  \bibfield  {author} {\bibinfo {author} {\bibfnamefont {S.}~\bibnamefont
  {Emori}},\ }\href@noop {} {\bibfield  {journal} {\bibinfo  {journal} {Nat.
  Mater.}\ }\textbf {\bibinfo {volume} {12}},\ \bibinfo {pages} {611} (\bibinfo
  {year} {2013})}\BibitemShut {NoStop}%
\bibitem [{\citenamefont {Lo~Conte}\ \emph {et~al.}(2015)\citenamefont
  {Lo~Conte}, \citenamefont {Martinez}, \citenamefont {Hrabec}, \citenamefont
  {Lamperti}, \citenamefont {Schulz}, \citenamefont {Nasi}, \citenamefont
  {Lazzarini}, \citenamefont {Mantovan}, \citenamefont {Maccherozzi},
  \citenamefont {Dhesi}, \citenamefont {Ocker}, \citenamefont {Marrows},
  \citenamefont {Moore},\ and\ \citenamefont {Klaui}}]{lo_conte}%
  \BibitemOpen
  \bibfield  {author} {\bibinfo {author} {\bibfnamefont {R.}~\bibnamefont
  {Lo~Conte}}, \bibinfo {author} {\bibfnamefont {E.}~\bibnamefont {Martinez}},
  \bibinfo {author} {\bibfnamefont {A.}~\bibnamefont {Hrabec}}, \bibinfo
  {author} {\bibfnamefont {A.}~\bibnamefont {Lamperti}}, \bibinfo {author}
  {\bibfnamefont {T.}~\bibnamefont {Schulz}}, \bibinfo {author} {\bibfnamefont
  {L.}~\bibnamefont {Nasi}}, \bibinfo {author} {\bibfnamefont {L.}~\bibnamefont
  {Lazzarini}}, \bibinfo {author} {\bibfnamefont {R.}~\bibnamefont {Mantovan}},
  \bibinfo {author} {\bibfnamefont {F.}~\bibnamefont {Maccherozzi}}, \bibinfo
  {author} {\bibfnamefont {S.~S.}\ \bibnamefont {Dhesi}}, \bibinfo {author}
  {\bibfnamefont {B.}~\bibnamefont {Ocker}}, \bibinfo {author} {\bibfnamefont
  {C.~H.}\ \bibnamefont {Marrows}}, \bibinfo {author} {\bibfnamefont {T.~A.}\
  \bibnamefont {Moore}}, \ and\ \bibinfo {author} {\bibfnamefont
  {M.}~\bibnamefont {Klaui}},\ }\href@noop {} {\bibfield  {journal} {\bibinfo
  {journal} {Phys. Rev. B}\ }\textbf {\bibinfo {volume} {91}},\ \bibinfo
  {pages} {014433} (\bibinfo {year} {2015})}\BibitemShut {NoStop}%
\end{thebibliography}%


\end{document}